\documentstyle[epsfig]{mn}
\def\apj{{ApJ.}}

\def\aap{{A\&A}}
\def\aj{{AJ}}

\begin{document}


\title[Observations of HDF South with {\em ISO} - I.]
{Observations of the {\em Hubble Deep Field South} 
with the {\em Infrared Space Observatory} - I. Observations,
data reduction and mid--infrared source counts}
 \author[Seb Oliver  {\it et al.}]
{ 
Seb Oliver$^{1,2}$, 
Robert G. Mann$^{1,3}$,
Ruth Carballo$^{4,5}$, 
Alberto Franceschini$^6$, 
\newauthor
Michael Rowan--Robinson$^1$, 
Maria Kontizas$^7$, 
Anastasios Dapergolas$^8$, 
\newauthor
Evanghelos Kontizas$^8$, 
Aprajita Verma$^1$, 
David Elbaz$^9$, 
Gian Luigi Granato$^6$,
\newauthor
Laura Silva$^{10}$, 
Dimitra Rigopoulou$^{11}$,
J. Ignacio Gonzalez--Serrano$^{5}$,
\newauthor
Steve Serjeant$^{1,12}$, 
Andreas Efstathiou$^1$,
Paul P. van der Werf$^{13}$
\\
$^1$Astrophysics Group, Imperial College London, Blackett Laboratory,
Prince Consort Road, London SW7 2BZ\\ 
$^2$Astronomy Centre, School of Chemistry, Physics and Environmental
Science, University of Sussex, Falmer, Brighton, BN1 9QJ\\
$^3$Institute for Astronomy, University of Edinburgh, Royal Observatory, Blackford Hill, Edinburgh, EH9 3NJ\\
$^4$Departamento de Matematica Aplicada y CC, Universidad de Cantabria, Avda. Los
Castros s/n, 39005 Santander, Spain\\
$^5$Instituto de F\'{\i}sica de Cantabria (CSIC-UC), Avda. Los Castros s/n,
39005 Santander, Spain.\\
$^6$Dipartimento di Astronomia --- Universita' di Padova, Vicolo 
dell'Osservatorio 5, I-35122, Padova, Italy\\
$^7$Department of Physics, 
University of Athens, Panepistimiopolis, GR-15783, Zografos, Greece\\
$^8$Astronomical Institute, National Observatory of Athens, Lofos Nymfon, 
Thission, P.O. Box 20048, 11810 Athens, Greece\\    
$^9$DSM/DAPNIA/SAp, CE - Saclay, Orme des Merisiers - Bat 709, 91191 
Gif-sur-Yvette Cedex, France\\
$^{10}$Astrophysics Sector, SISSA, Via Beirut 2-4, 34013 Trieste, Italy\\ 
$^{11}$Max-Planck-Institut f\"ur extraterrestrische Physik, Postfach
1603,85740 Garching, Germany\\
$^{12}$Unit for Space Sciences and Astrophysics, School of Physical Sciences,
University of Kent, Canterbury, Kent, CT2 7NZ\\
$^{13}$Leiden Observatory, P.O. Box 9513,  NL-2300 RA Leiden, The
Netherlands
}

\pagerange{\pageref{firstpage}--\pageref{lastpage}}
\pubyear{2002}
\volume{}

\label{firstpage}

\maketitle
\begin{abstract}
We present results from  a deep mid--infrared survey of the Hubble Deep Field South (HDF--S) region 
performed at 7 and 15$\mu$m with the CAM instrument on board the {\em
Infrared Space Observatory} ({\em ISO}).
The final map in each band was constructed by the coaddition of four independent rasters, registered using 
bright sources securely detected in all rasters, with the absolute astrometry being defined by
a radio source detected at both 7 and 15$\mu$m. We sought
detections of bright sources in a circular region of radius 2.5 arcmin at the centre of
each map, in a manner that simulations indicated would produce highly reliable and complete source 
catalogues using simple selection criteria. Merging source lists in the two bands yielded a 
catalogue of 35 distinct sources, which we calibrated photometrically using photospheric models of
late--type stars detected in our data. We present extragalactic source count results in both bands,
and discuss the constraints they impose on models of galaxy evolution models, given the volume of
space sampled by this galaxy population.\\
\end{abstract}

\section{Introduction}\label{intro}

\begin{table*}
\caption{{\em ISO} Observation Log. This table gives some details from
the {\em ISO} databases for each of the {\em ISO} HDF--S observations:
The target name, coordinates,  Observation Number (OSN), the time
spent on target in seconds (TDT), the revolution number (REV), the status and the
date.  Note that two observations (OSN 4 and 8) failed, but were
repeated
on 27 and 29 November.}
\begin{tabular}{ccccccrr}
TARGET       &      RA (J2000)  &  DEC (J2000)   & OSN &  TDT & REV
&STATUS     &    Date\\~\\ \hline	   
HDF-1 LW2    &    22h 32m 57.5s &-60d 33' 10.0"  &  1 &  7825 &702 &Observed   &     17 Oct 1997 \\
HDF-4 LW2    &    22h 32m 53.9s &-60d 33' 00.0"  &  7 &  7825 &702 &Observed   &     17 Oct 1997 \\
HDF-2 LW2    &    22h 32m 56.4s &-60d 32' 51.8"  &  3 &  7825 &704 &Observed   &     19 Oct 1997 \\
HDF-4 LW3    &    22h 32m 53.9s &-60d 33' 00.0"  &  8 &  7497 &722 &Failed     &      6 Nov 1997 \\
HDF-2 LW3    &    22h 32m 56.4s &-60d 32' 51.8"  &  4 &  7497 &722 &Failed     &      6 Nov 1997 \\
HDF-3 LW2    &    22h 32m 55.0s &-60d 33' 18.2"  &  5 &  7825 &723 &Observed   &      7 Nov 1997 \\
HDF-3 LW3    &    22h 32m 55.0s &-60d 33' 18.2"  &  6 &  7497 &723 &Observed   &      7 Nov 1997 \\
HDF-1 LW3    &    22h 32m 57.5s &-60d 33' 10.0"  &  2 &  7497 &723 &Observed   &      8 Nov 1997 \\
HDF-4 LW3    &    22h 32m 53.9s &-60d 33' 00.0"  &  8 &  7497 &742 &Observed   &     27 Nov 1997 \\
HDF-2 LW3    &    22h 32m 56.4s &-60d 32' 51.8"  &  4 &  7497 &745 &Observed   &     29 Nov 1997 \\

\end{tabular}

\end{table*}

One of the most notable achievements of the {\em Hubble Space
Telescope} ({\em HST}) has been to lead and inspire the concerted
multi--wavelength programme of observations of the {\em Hubble Deep
Field} (HDF, Williams et al. 1996)\nocite{Williams et al. 1996} 
region. As part of that
campaign we observed the HDF at 6.7 and 15 $\mu$m using the
{\em ISO}--CAM instrument (Cesarsky et al. 1996) on the {\em
Infrared Space Observatory} ({\em ISO}: Kessler et al. 1996). From
the maps that resulted from these observations (Serjeant et al. 1997)
we extracted sources in both bands (Goldschmidt et al. 1997), whose
number counts implied a strongly--evolving population of starburst
galaxies (Oliver et al. 1997).  Following the association of these
sources with galaxies in optical HDF catalogues (Mann et
al. 1997) we derived an infrared luminosity density that suggested a
higher star--formation rate in the HDF region than indicated
by optical studies (Rowan-Robinson et al. 1997): the importance of 
dust obscuration in estimating the star-formation rate
has been confirmed by other {\em ISO} surveys e.g. 
\cite{Flores et al. 1999}, from detailed consideration of the optical
measures of star formation \cite{Steidel et al. 1999} and from 
the intercomparison of different star formation indices 
\cite{Cram et al. 1998}

Difficulties with the {\em ISO}
6.7 $\mu$m data led us to re--observe the HDF at that
wavelength.  These new data, together with a consensus view of the
interpretation of our {\em ISO} HDF--N data derived from the combined
experience of the several groups that re--analyzed them (Aussel et
al. 1999, Desert et al. 1999) in the light of developing knowledge of
the properties {\em ISO}--CAM data will be the topic a subsequent
paper, as will a revised and updated scientific interpretation of the
{\em ISO HDF} data.

Following the success of the  HDF project, a similar programme
of {\em HST} observations was planned for the southern hemisphere,
and the region of the Hubble Deep Field South (HDF--S)
has become the target of a similarly wide--ranging multi--wavelength 
programme\footnote{Details of the HDF--S programme can be found
at \verb+http://www.stsci.edu/ftp/science/hdf/hdfsouth/hdfs.html+} of 
observations. This paper describes our contribution to that project,
through our mapping of the HDF--S with {\em ISO}--CAM. We
mapped the HDF--S WFPC2 fields at both 6.7 and 15 $\mu$m, as in
the northern HDF, but with a slightly different observational 
strategy (described in Section 2), motivated by our experience with the 
{\em ISO HDF} data. Section 3 describes our data reduction
procedures, and Sections 4 and 5 source extraction and photometric calibration,
respectively. In Section 6 we describe simulations of the data
performed to facilitate assessment of the reliability of the source
catalogues we present in Section 7 and to compute the effective area
of the survey as a function of flux cut, as this is required for
computation of the source counts, which is the topic of Section 8. 
Finally, Section 9 presents a discussion of the results of this
paper and the conclusions we draw from them. In an accompanying
paper (Mann et al., 2002, hereafter Paper II) we seek associations
for these sources in optical/near--infrared and radio surveys of
the HDF--S region, and present star formation rate estimates for the
sources for which we find associations.

\section{The Observations}\label{obs}

The team undertaking the European Large Area {\em ISO} Survey
(ELAIS\footnote{For details see the ELAIS Home Page: 
\verb+http://astro.ic.ac.uk/elais+}, Oliver et al. 2000) were awarded 61.3 ks  to observe
the HDF--S in a successful application to the {\em ISO
Supplemental Call}.  The observations were
carried out using the {\em ISO}--CAM instrument (Cesarsky et al.
1996) between 17 October and 29 November 1997.  In the light 
of an increased understanding of the properties of {\em ISO}--CAM data,
and of the galaxy population they probe, gained by our own
{\em ISO} HDF observations and from other deep {\em
ISO}--CAM surveys (e.g. Taniguchi et al. 1997, Elbaz et
al. 1999\nocite{Elbaz et al. 1999}) 
we made some alterations to the observing strategy that we
used for the {\em ISO} HDF, as  reported by Serjeant et al. (1997).

\begin{table}
\caption{Observation parameters  for the ISO HDF South}\label{tab:aots}
\begin{tabular}{lll}

Parameter       & \multicolumn{2}{c}{OSN}\\ 
                & 1,3,5,7  & 2,4,6,8 \\ \hline
Filter          & LW2      & LW3 \\
Band centre [$\mu$m]  & 6.7      & 15  \\
{\em Gain}            & 2        & 2 \\
{\em Tint} [s]          & 10       & 5 \\
{\em NEXP}            & 10       & 20 \\
{\em NSTAB}           & 80       & 80 \\
{\em PFOV} [arcsec]    & 6        & 6\\
$M,N$           & 8,8      & 8,8 \\
$dM,dN$ [arcsec] & 27       & 27 \\

\end{tabular}

\end{table}

\begin{figure}
\hspace{3cm}
\epsfig{file=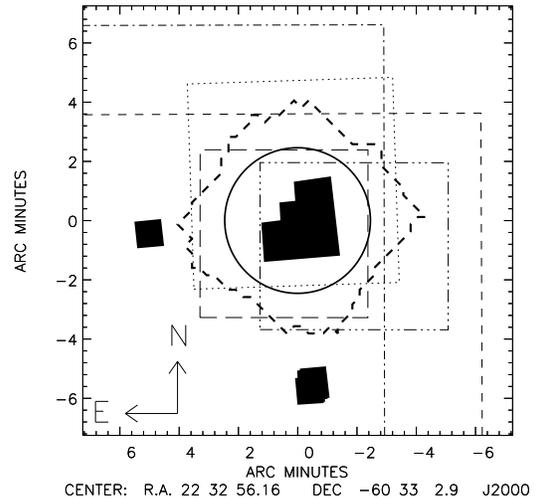,angle=0,width=10cm}
\caption{This figure shows the location of our ISO rasters with
respect to those of other datasets taken in the area. The shaded
regions mark the {\em HST} fields, with the STIS and NICMOS fields
to the east and south of the WFPC2 field, respectively. The 
thick--dashed irregular shape and the solid
circle show, respectively, the maximum extent of our ISO
coverage (the coverage in the two bands differs slightly)
 and the region from which the source catalogues of Section
\ref{catalogues} were selected. The remaining lines show boundaries of
four optical/near-IR surveys discussed in Paper II,
as follows: (i) dotted line -- AAT prime focus imaging survey of Verma
et al.  (2002); (ii) dashed line -- CTIO BTC survey of Gardner et
al. (1999); (iii) dot-dashed line -- CTIO BTC survey of \protect\cite{Walker 1999};
(iv) dot-dot-dot-dashed line -- ESO EIS optical imaging survey of da
Costa et al. (1998); and (v) long dashed line -- ESO EIS near-infrared
survey of da Costa et al. (1998).}
\label{fig:fields}
\end{figure}

As listed in Table 1, eight rasters were taken, one with the LW2 (6.7
$\mu$m) and one with the LW3 (15 $\mu$m) filter  at each of four positions, 
with raster centres offset by fractional pixel widths to improve the spatial
resolution of the final map obtained by their coaddition. The 
parameters used for these observations are listed in Table \ref{tab:aots}. 
The values for {\em Gain}, {\em Tint} (the integration time per
readout), {\em NEXP} (the number of readouts per
pointing), and {\em NSTAB} (the number of readouts allowed for
stabilization prior to
the raster) remained unchanged, as those used by Serjeant et
al. (1997) still appeared to be optimal for
the required depth.  The pixel field of view, {\em PFOV},  at
6.7$\mu$m was changed from 3$\arcsec$ to 6$\arcsec$ (matching that used at
15 $\mu$m), since the 
{\em ISO--HDF} images at 6.7$\mu$m were not confusion limited, and the
improvement in signal--to--noise ratio, and areal coverage,
 obtained by moving to larger
pixels was expected to outweigh
the loss in resolution.  The other significant change was to increase
the raster step size, which had two effects. Firstly, with the step size
now larger than the point spread function (PSF), consecutive
pointings would no longer have significantly correlated signal, making
the removal of noise which was correlated in time (e.g. $1/f$ noise) 
much easier. Secondly, the larger
pixel area meant that the full survey area could be covered by each
individual $8 \times 8$ raster, with the complete survey being made up
by stacking the four independent rasters in each passband. This
contrasts  with the technique adopted in the 
{\em ISO}--HDF of partially overlapping deeper rasters, and has 
several advantages: it reduces correlated noise problems; readily provides
direct assessment of source reliability, through looking for
detections in the independent
maps;  and facilitates the registration of the maps, through the presence of a
greater number of bright sources in each raster (helped further by the 
lower Galactic latitude of the {HDF--S}, providing more bright stellar sources).
In Figure \ref{fig:fields} we show the location of our ISO rasters with
respect to those of various optical/near-IR datasets taken in the 
{HDF--S} area: this illustrates that while none of these surveys
covers the whole of the area we mapped with ISO, the region from which
we select sources in Section \ref{catalogues} is covered, at least
partially, by several imaging surveys in different wavebands, as 
discussed in more detail in Paper II.

\section{Data Reduction}\label{reduction}

Similarly to the {\em ISO} HDF data, and in contrast to the ELAIS
data (Oliver et al. 2000),
we do not expect to detect many sources in the signal from a
single pixel as it scans across the sky (the ``time-line'').  
Most sources will only be detected when all the
overlapping scans are co-added.  The data reduction thus proceeds by
filtering each time-line for artifacts and then combining these to produce a map for
each raster.  These raster maps are then co-aligned and co-added to
produce a single map from which sources can be extracted.
Most of the data reduction described in these sections was carried out
using the Interactive Data Language (IDL
\footnote{see {\tt www.rsinc.com}}), with  some steps done using the
{\em ISO}--CAM Interactive Analysis (CIA: Ott et al. 1998) software.

\subsection{Time-series Filtering}

The first stage in the data reduction treats the scan of each pixel
across the sky independently.  This time--series is filtered to reduce
the impact of noise features and optimize the signal at each static
pointing.  At this stage the individual pixel responses are also
estimated using a Gaussian fit to the scan to determine a sky flat-field
correction which is normalized to the median from the central pixels,
as in Serjeant et al. (1997).

\begin{figure}
\epsfig{file=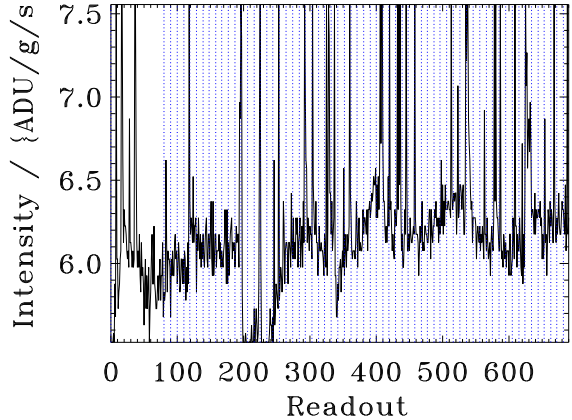,angle=90,width=8cm}
\caption{Example pixel history.  Minimum and maximum levels are
adjusted to exclude extreme outliers.  The positions of slews
are marked with vertical dotted lines.  This particular pixel
is row 16, column 16 from the 6.7$\mu$m observation HDF-1 }\label{fig:egpixel}
\end{figure}

The original data reduction of the {\em ISO}--HDF applied a simple
threshold filtering of very short time-scale features (cosmic ray
hits), and we apply the same method to these data. However, for the
original reduction of the {\em ISO}--HDF data we anticipated that there
might be significant source confusion, leading to real structure in the sky
background, and so we did not apply any filtering for low frequency
noise.  At 6.7 $\mu$m the {\em ISO}--HDF was not significantly confused and
the revised observing strategy we used for the HDF--S
reduces any correlated signal between
successive pointings.  We thus decided to adopt a more aggressive
filtering strategy for these data.  
It can be seen from Figure \ref{fig:egpixel} that
there is significant correlated noise at a variety of time-scales.  The
filtering technique we adopt is similar to that
used by Desert et al. (1999).  We subtract a time variable 
background level from all the readouts in a time-line.
The background level is estimated for each pointing, being the average of
the readouts in the two previous and the two subsequent pointings.

This filtering scheme will go awry where sources lie in the pixels
used for background estimation, so we perform a second iteration.
We mask out the bright sources detected from the first pass
(see Section~\ref{sources}) and then an
additional filter is applied to exclude readouts not already flagged
as
sources that deviate from the
estimated background by more than 5$\sigma$.
This procedure will not affect
remaining sources which would  be at a very low level of significance in
single pointings. After applying these filters the mean of the readouts
is calculated over each pointing, and the resulting noise statistics are
summarized in Table \ref{tab:noise}.

\begin{table*}

\caption{
Statistical properties of the reduced data sets.  All units are
instrumental (ADU/g/s/pixel). 
The mode is taken from all valid readouts and 
pixels.  Fluctuations are estimated
by fitting a Gaussian to the distribution (first lines) and by
calculating the RMS (second lines).
The difference between these two measures gives some idea of the
non-Gaussianity. Fluctuations are quoted for the readout and 
pointing time scales (third and fourth column) 
and are estimated from the time line of one pixel near the centre 
of the image. 
In the filtered readouts the RMS
fluctuations are similar to the $\sigma$ estimated from a Gaussian
fitting, suggesting that the noise is reasonably Gaussian. If the
fluctuations are white noise they should reduce by a factor of 
$\sqrt{\rm NEXP}$ (i.e. $\sqrt{10}$ at 6.7 $\mu$m and 
$\sqrt{20}$ at 15  $\mu$m) when averaging over a pointing.
The fluctuations over an entire detector image (after coaddition of
all readouts in a pointing) are quoted in column 5 and
estimated from all valid pixels and images.  The noise in the images is calculated separately for regions
with different numbers of pointings and is fitted as $\sqrt{\rm
NPOINT}$, the values quoted in the Table are for ${\rm NPOINT}=1$,
and for our maps  $\overline{{\rm NPOINT}}\sim 13$.
Finally, we compute the noise statistics for one of our mis--mosaiced 
maps (as described in Section \protect\ref{simulations}), column 6. Here we only
quote the $\sigma$ values as the RMS measures are more strongly
influenced by the residual source signals in parts of the images with
small values of ${\rm NPOINT}$.
}\label{tab:noise}

\begin{tabular}{lrrrrll}
Observation& Mode &  \multicolumn{2}{c}{Readout} & \multicolumn{1}{c}{Point.} & \multicolumn{1}{c}{Image} & \multicolumn{1}{c}{Mis--}\\ 
           &       & \multicolumn{1}{c}{Raw}  & \multicolumn{1}{c}{Filt.} &   & &\multicolumn{1}{c}{mosaiced}\\ \hline
 HDF-1 LW2 &  6.30 & 0.21 & 0.14 & 0.08	 & 0.089 & 0.088\\ 
           &       & 2.98 & 0.14 & 0.03  & 0.22\\ 
 HDF-2 LW2 &  6.47 & 0.13 & 0.11 & 0.06  & 0.081 & 0.079\\ 
           &       & 1.90 & 0.11 & 0.01  & 0.25\\ 
 HDF-3 LW2 &  7.38 & 0.43 & 0.19 & 0.19	 & 0.11  & 0.10\\ 
           &       & 3.88 & 0.21 & 0.07  & 0.20\\ 
 HDF-4 LW2 &  6.37 & 0.20 & 0.13 & 0.08	 & 0.083 & 0.083\\ 
           &       & 2.32 & 0.14 & 0.04  & 0.22\\ 
           & 	   &	  &	 &	 &\\ 
 HDF-1 LW3 & 33.49 &      & 0.33 & 0.13	 &0.22   & 0.25\\ 
           &       &      & 0.32 & 0.06	 &0.29\\ 
 HDF-2 LW3 & 37.39 &      & 0.30 & 0.12  &0.24   & 0.26\\              
           &       &      & 0.29 & 0.05	 &0.30\\ 
 HDF-3 LW3 & 34.04 &      & 0.34 & 0.18  &0.23   & 0.25\\     
           &       &      & 0.34 & 0.14	 &0.30\\
 HDF-4 LW3 & 35.09 &      & 1.53 & 1.50  &1.28   & 1.28\\ 
           &       &      & 1.23 & 2.80  &1.50\\ 
\end{tabular}

\end{table*}

\subsection{Mosaicing Independent Rasters}

The detector image at each raster position needs to be projected onto
the sky.  For this process we use a ``shift-and-add'' technique. We
generate a blank sky map with 1\arcsec pixels.  Then, for each raster
pointing we determine which sky pixels lie within the geometrical footprint of
each detector pixel.  
In doing this we take into account the field
distortions (Aussel et al. 1999) in the {\em ISO}--CAM data: {\em N.B.}
it was not possible to take these into account in our original 
{\em ISO}--HDF--N  reductions since the distortions had not been well
characterized at that time.
The average intensity of all detectors covering a sky pixel is
calculated using a number of estimators.  It was found that the
median produced the smallest fluctuations in the resulting
sky maps  (suggesting that some residual non-Gaussian noise was
present), and so we adopted this estimator. The use of the 
geometrical footprint was a somewhat arbitrary choice and not 
the optimal one for point sources; since in the extraction of point sources
we convolve the image with another kernel we have broadened
the effective point spread function by using this.

\begin{figure}
\epsfig{file=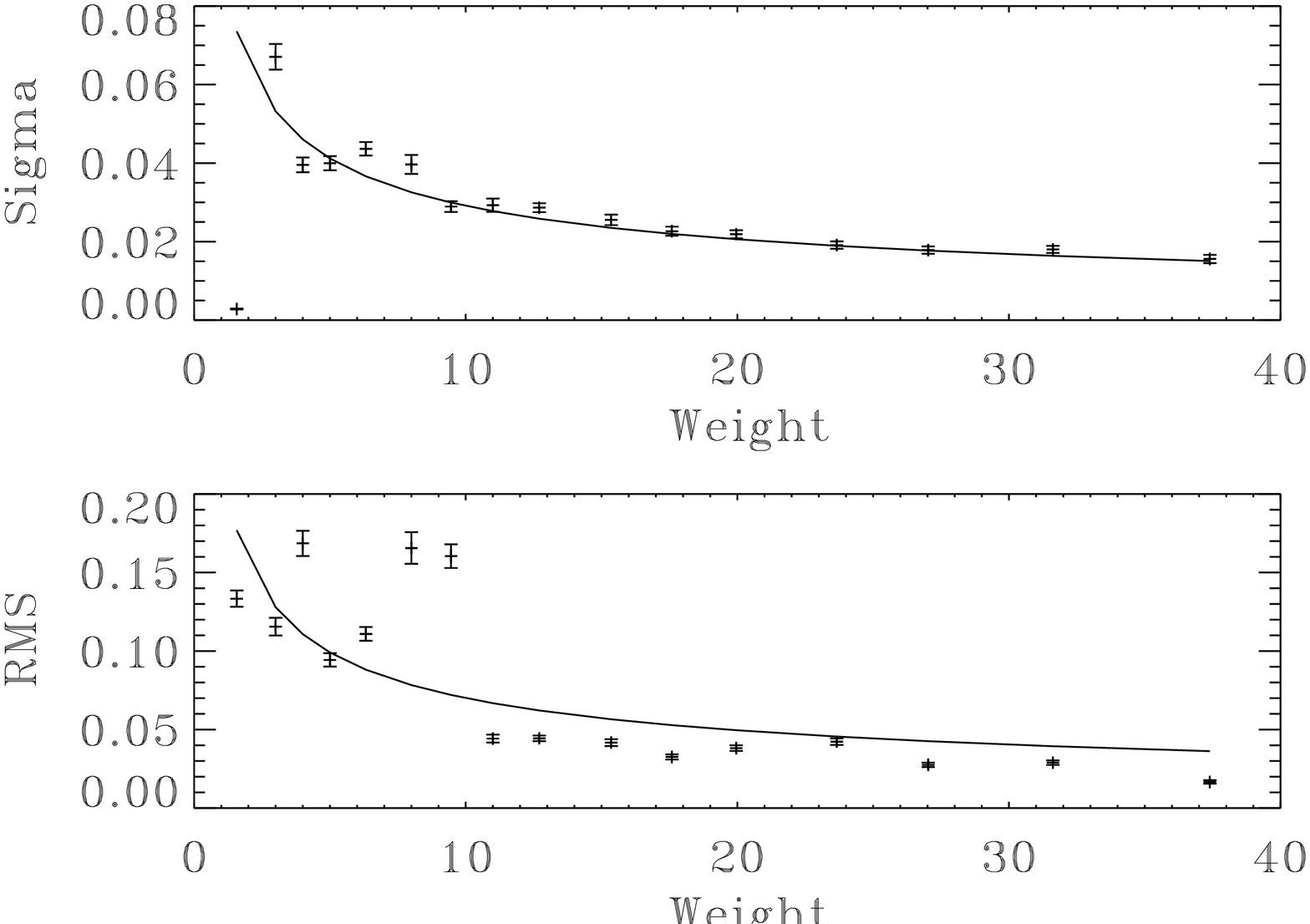,angle=90,width=8cm}
\caption{Noise estimates as a function of number of pointings 
($NPOINT$, labeled as ``Weight''), for the
LW2 (6.7$\mu$m)  HDF-1 observartions.
Upper panel $\sigma$ from a Gaussian fit, lower panel
$RMS$.  All units are ADU/g/s/pixel}\label{fig:noise1}
\end{figure}

\begin{figure}
\epsfig{file=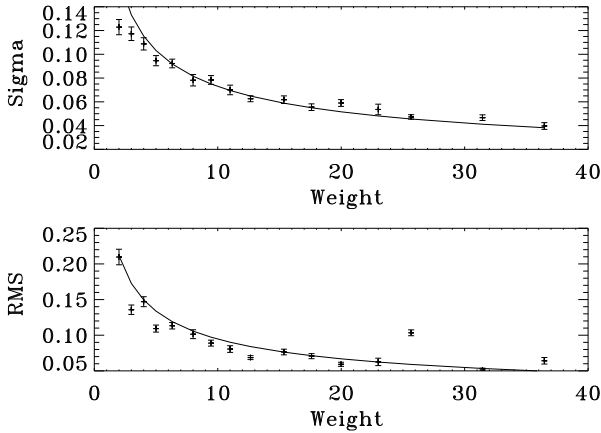,angle=90,width=8cm}
\caption{Noise estimates as a function of number of pointings 
($NPOINT$, labeled as ``Weight'')for the
LW3 (15$\mu$m) HDF-1 observations.
Upper panel $\sigma$ from a Gaussian fit, lower panel
$RMS$.  All units are ADU/g/s/pixel}\label{fig:noise2}
\end{figure}

The noise in these maps is estimated by constructing a histogram
of pixel values and fitting this with a Gaussian, as well as
computing directly an RMS.   However, since
the number of independent pointings varies as a function of sky
position, we calculate these statistics for regions with similar
numbers of pointings.  The results from these assessments indicate that
noise reduces as expected for independent pointings (i.e. the
residual noise-correlations between pixels are at a very low level: see Figs.
\ref{fig:noise1} and  \ref{fig:noise2}). The RMS
fluctuations in the maps are larger than the $\sigma$ estimated from
the Gaussian fitting, indicating non-Gaussian fluctuations, which
we attribute to real sources, either distinct or confused.

The noise can also be investigated by corrupting the astrometry
information and thereby diluting the signal from real sources.
We describe this ``mis--mosaicing'' technique in more detail in 
Section \ref{simulations}.  The noise statistics from a typical
mis--mosaiced field, together with those from the final images, are
also summarized in Table \ref{tab:noise}.  The $\sigma$ values 
from the mis--mosaiced fields are very similar to the real fields
confirming that these are reasonably representative of the noise.
The RMS values for the mis--mosaiced fields were invariably higher
but this was because the RMS in regions of lower coverage were
higher, since real sources could not be properly filtered out in 
these regions:  this is consistent with  our hypothesis
that much of any residual non--Gaussian noise is due to real sources.

\subsection{Raster registration}

Since the survey strategy means that each independent raster covers
approximately the same relatively large area (c.f. {\em ISO}--HDF) a
number of bright sources are clearly visible in each raster.  For each
band we thus selected a number of these sources which had good 
signal--to--noise and which were not located close to the edges of the map (where
the noise is less well behaved and the field distortions are more
significant) for use in registering the four maps.  
The positions of these sources are denoted by 
$(x_{i,j},y_{i,j})$, where the $i$ subscript labels the source and $j$
subscript labels the map. We then computed the mean $(\overline{x_i},\overline{y_i})$ 
position of each source across all four independent maps, weighted
by $w_j$, the  mean SNR of that map (estimated from the mean SNR in all
the sources).  For each map we then estimated a mean offset
$\delta x_j=\sum w_i(x-\overline{x_i})/\sum w_i$, where the weight for each
source $w_i$ is estimated from the mean SNR for that source over all
maps. This process does not require any assumptions about the
relationship between the {\em ISO} sources and sources detected in any
other wavebands, and is also likely to be more robust than the
cross--correlation of the full image (including the noisy regions)
with, for example, a radio map of the same field.
The mean offsets were rounded to
the nearest pixel (1\arcsec). The overall astrometric reference frame
was defined later, see Section 7 below.

\begin{figure*}
\epsfig{file=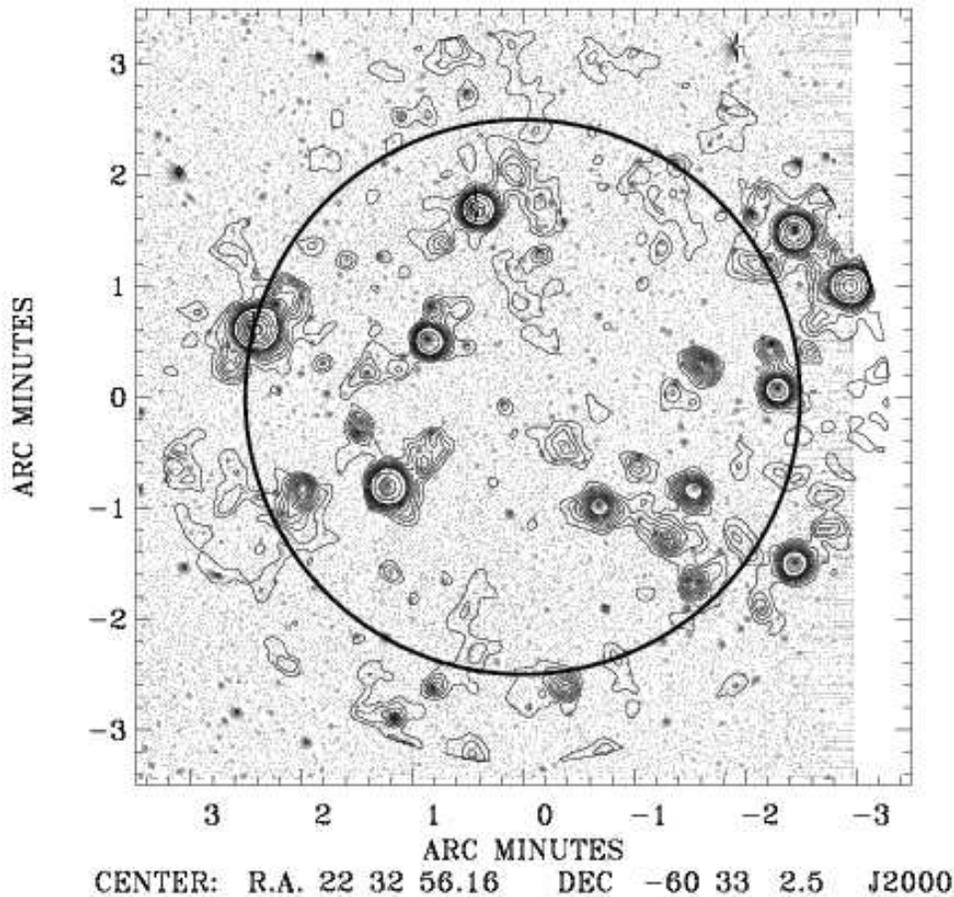,width=18cm}
\caption{LW2 (6.7$\mu$m) signal--to--noise map.  
This figure plots contours in the LW2 signal--to--noise map after it has been
smoothed with the point-source-detection kernel.  The lowest contour
level has signal--to--noise = 1 and subsequent intervals are 1 until
signal--to--noise = 10 where after the contours are logarithmically
spaced. {\em ISO} data is plotted out to a radius of  3.3\arcmin\.  The background image is the CTIO BTC survey of \protect\cite{Walker 1999}.  The circle indicates the 2.5\arcmin\ boundary of the region within
which we extracted sources for our catalogues. Overlays of subsections
of the data onto colour HST images are available from our WWW page ({\tt astro.ic.ac.uk/hdfs}).}\label{fig:lw2snr}
\end{figure*}

\begin{figure*}
\epsfig{file=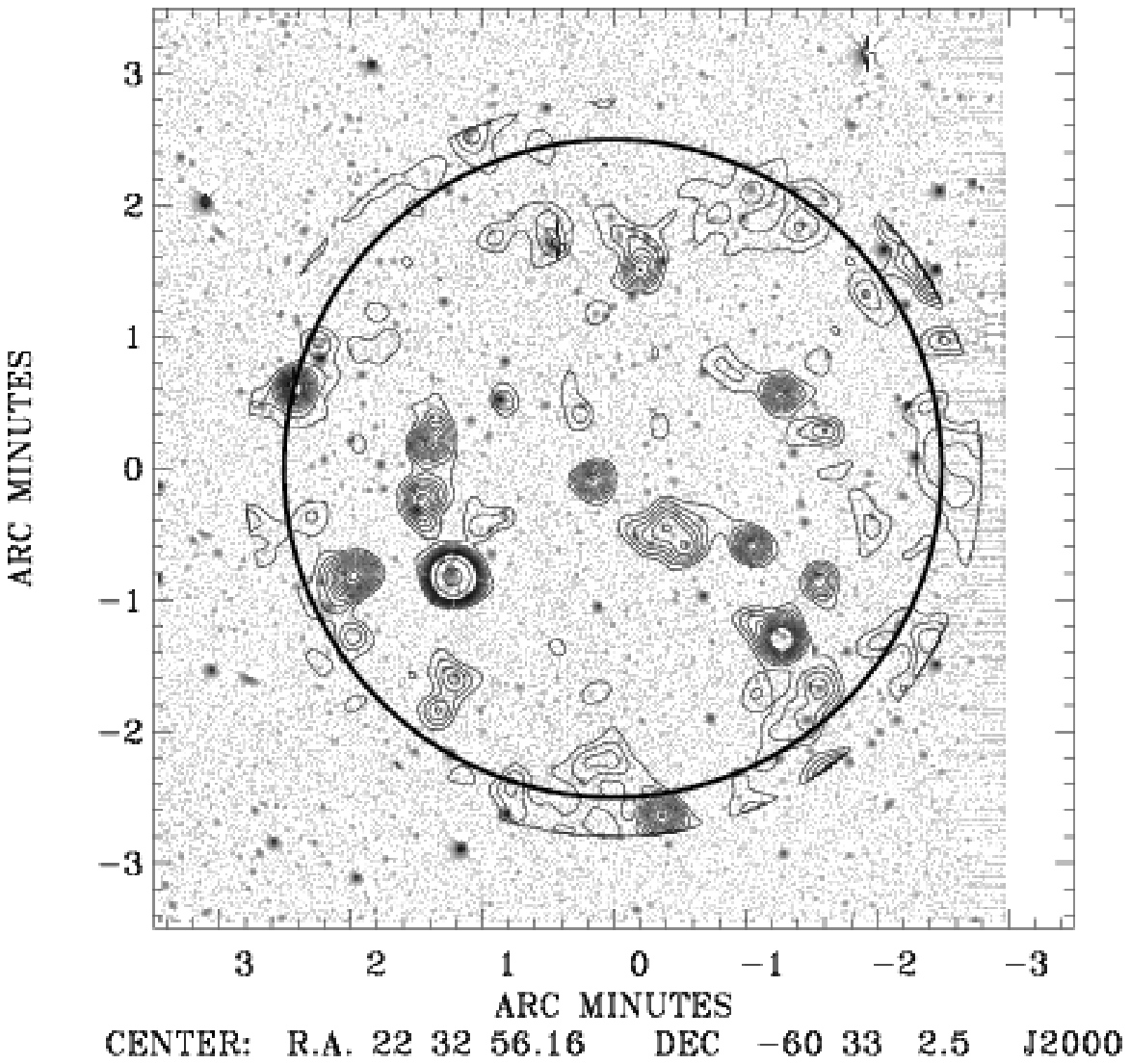,width=18cm}
\caption{LW3 (15$\mu$m) signal--to--noise map.  
This figure plots contours in the LW3 signal--to--noise map after it has been
smoothed with the point-source-detection kernel.  The lowest contour
level has signal--to--noise = 1 and subsequent intervals are 1 until
signal--to--noise = 10 where after the contours are logarithmically
spaced.   {\em ISO} data is plotted out to a radius of  2.8\arcmin\.  The background image is the CTIO BTC survey of \protect\cite{Walker 1999}. The circle indicates the 2.5\arcmin\ boundary of the region within
which we extracted sources for our catalogues. Overlays of subsections
of the data onto colour HST images are available from our WWW page ({\tt astro.ic.ac.uk/hdfs}).}\label{fig:lw3snr}
\end{figure*}

The registered images were then co-added using an
inverse variance weighting. The variance estimated was proportional to
the number of pointings within a raster and scaled using the value for
$\sigma/\sqrt{\rm NPOINT}$ estimated from the Gaussian fitting described
above.  The resulting maps had some small residual background. This background 
(which was not always positive) should in principle have been removed by the time-line filtering, although
some residual would be expected from an overall gradient in the
time-lines.  In any case, the background  was estimated from the mean of 
the Gaussian fitted to the histogram of 
the map pixel values, which was a good estimate of the mode.
The resulting co-added signal--to--noise maps at 6.7 and 15 $\mu$m are presented in Figures
\ref{fig:lw2snr} and \ref{fig:lw3snr}, respectively.

\section{Source Detection}\label{sources}

We expect most of the sources in these maps to be point sources, so it is
 more appropriate to detect sources by a convolution technique
rather than to use a connected pixel algorithm.  We also expect to be
close to the confusion limit, so the choice of smoothing kernel is
important:  where the signal is dominated by a single source the
optimal kernel is the point-spread function (PSF), while, 
for confused images, the
likely presence of other sources in the wings of the PSF
 will make this kernel non-optimal.  The solution is either to
truncate the PSF kernel at some appropriate distance
or to use a narrower kernel. 
Theoretically, the PSF should be that of the Airy disk
defined by the telescope aperture, convolved with the square pixel
aperture of 6$\arcsec$, and the {\em CIA} calibration PSFs are
similar to this.  In practice, however, the PSF will be broadened
by our mapping footprint, and any inaccuracies in registration and/or
the field-distortion correction applied.  
The profiles we used for source extraction are Gaussian, with
a FWHM of 6$\arcsec$ and 10$\arcsec$ at 6.7$\mu$m and
15$\mu$m respectively, and both are truncated at a radius of 12$\arcsec$.
Table  \ref{tab:psfs} compares the FWHM values for these mode
 PSFs with those estimated from the sources detected in the data. It
 shows that, as desired, the model PSFs used for source 
detection, while being slightly larger than the theoretical PSF, are slightly
 narrower than the empirical PSF derived from the sources themselves.

\begin{table*}

\caption{ Full-width-half-maxima (FWHM, in arcsec) for theoretical and empirical 
point-spread functions: see text for details.  
The column marked ``used'' gives the FWHM values for the model PSFs
used for source extraction,  the ``star''
PSF is estimated from the brightest star in the LW2 image, and the ``sources''
PSF from the sources detected above 20$\sigma$ in each
image (8 sources for LW2 and 1 source for LW3). Empirical
FWHM are calculated by fitting a 2-D Gaussian to the observed PSF.}
\label{tab:psfs}

\begin{tabular}{lccccccc}
Filter & $\lambda[\mu$m]    & Airy disk + pixel & CIA &  Used & Star & Sources
       \\ \hline 
LW2    & 6.7              & 5.8       & 7.3 & 6.0 &  10.4 &   8.3 \\
LW3    & 15               & 7.3       & 9.0 & 10.0 &       &  10.4 \\
\end{tabular}

\end{table*}

An initial candidate list of sources was selected in each band, comprising peaks in
the respective convolved, co-added signal-to-noise ($SNR0$) map above $3\sigma$.  
For each of
these peaks we returned to the individual raster maps and computed a
number of additional statistics.  These included the 
signal-to-noise ratio in each map ($SNRn,n=1-4$) and the number of pointings at the source
position in these maps ($NPOINTn,n=1-4$).  One  quantity derived ($PICK$) was the
number of detections with $SNRn>1$ and $NPOINTn>4$. We also computed the mean and RMS deviation of the
flux for the source over all maps, together with the ratio of these
($SNR5$).  Using the simulations described in
Section \ref{simulations} we could then assess and employ these various statistics to
define suitable simple criteria for filtering the candidate list to produce a
highly reliable source list with reasonable completeness.

\section{Photometric calibration}\label{calibration}

One difference between the {HDF South} and the {HDF North} is
that the former is located at lower Galactic latitude and so has a
higher proportion of stars.  This is extremely useful for the
calibration of the {\em ISO} data, since a number of the bright stars
are detected, and photospheric model spectra can be used to estimate
their 6.7 and 15 $\mu$m fluxes. Our calibration procedure was to 
identify a number of stars for which we could accurately predict
mid-infrared fluxes and then measure  their fluxes directly from
the {\em ISO} maps.  Since  the stars are known sources with known
positions we are
not concerned about the reliability of their detection, so these stars do not
necessarily appear in our source lists, and  their fluxes can be
fainter than the faintest sources in our complete samples.

To model accurately the stellar fluxes we ideally need spectral
classifications and accurate magnitudes for the stars. We 
inspected and classified the star spectra taken on the AAT
with the LDSS (Glazebrook {\em et al.}, in preparation: see  \verb+http://www.aao.gov.au/hdfs/Redshifts/+).
We supplemented this list with additional stars that had been 
detected in our {\em ISO} maps (prior to the filtering applied to produce 
a highly reliable source list presented in Section \ref{catalogues}). We 
cross-correlated all these objects
with the ESO optical catalogues (da Costa et al. 1998) and  the AAT
catalogues (Verma et al. 2002) to obtain optical and NIR magnitudes.
This combined sample of stars is listed in Table \ref{tab:stars}.

\begin{table*}
\caption{Stars in {HDF South}.  This is an inhomogeneous list of
stars collected for the process of calibration.  Spectral types were
determined from the AAT spectra.  Optical magnitudes are in the AB
system and come from the ESO survey (da Costa et al. 1998) or the AAT
survey (Verma et al. 2000).  Observed 6.7 and 15$\mu$m fluxes ($So$) are in raw
instrumental units. Fluxes are taken directly from the maps. One
flux was measured to be negative (with large error) and is quoted as
such,  since this would be valid in a calibration fit,
however this point has not been used in the calibration.
Predicted  6.7 and 15$\mu$m ($Sp/\mu$Jy) are estimated for a
few of these objects as described in the text.  Magnitudes quoted as
99.99 indicated saturated measurements}\label{tab:stars}
\begin{tiny}
\begin{tabular}{clrrrrrrrrrrrr} 
 RA,Dec (J2000)		  &  Type   & U    &  B    &  V  &   R  &  I   &  J   &   H  &  K   &
$So_{6.7}$ & 
$So_{15}$ &  
$Sp_{6.7}$& $Sp_{15}$ \\
22 32 31.94  -60 32 00.7 &         &      & 11.94 &     & 12.17&      &      &      &      &  77.42&    23.79&         &       \\
22 32 36.19  -60 34 31.5 &         & 21.56& 19.81 &18.14& 17.00& 15.80&      &      &      &  17.66&    10.79&    626.0&  134.0\\
22 32 36.23  -60 31 31.5 &         & 20.38& 18.51 &16.92& 15.80& 14.60&      &      &      &  42.17&    26.01&    600.0&  130.0\\
22 32 37.48  -60 32 57.3 &         & 21.30& 19.38 &17.73& 16.70& 15.70&      &      &      &  12.56&     6.33&    611.0&  134.0\\
22 32 39.41  -60 31 22.6 &  G0 V   & 18.02& 17.03 &16.38& 16.10& 16.00&      &      &      &   2.17&     7.23&         &       \\
22 32 39.88  -60 33 23.6 &  M2.5 V & 99.99& 99.99 &99.99& 99.99& 26.40&      &      &      &   0.40&     5.38&         &       \\
22 32 40.70  -60 33 24.1 &         &      &       &     &      &      &      &      &      &   0.39&     5.30&         &       \\
22 32 43.51  -60 33 51.0 &         & 20.51& 20.11 &20.23& 20.01& 19.85& 19.78& 19.42& 19.53&   4.35&    10.76&         &       \\
22 32 47.45  -60 32 00.0 &  M2.5 V & 99.99& 24.05 &22.59& 21.31& 19.80& 18.36& 18.22& 18.42&   0.45&    -1.17&         &       \\
22 32 50.50  -60 34 00.8 &  M2 V   & 21.98& 20.24 &18.71& 17.90& 17.13& 16.41& 16.21& 16.49&   3.74&     0.96&         &       \\
22 32 50.62  -60 34 04.0 &         & 21.90& 20.23 &18.70& 17.84& 17.01& 16.42& 16.21& 16.49&   3.74&     1.30&    166.0&   36.5\\
22 32 54.90  -60 31 44.1 &         & 23.41& 21.70 &20.18& 19.30& 18.32& 17.51& 17.26& 17.51&   0.92&     6.34&         &       \\
22 32 56.72  -60 35 49.5 &         & 27.63& 25.81 &24.35& 23.50& 22.90&      &      &      &   1.00&     9.69&         &       \\
22 32 59.50  -60 31 19.2 &  G2 III & 15.11& 14.63 &13.91& 14.14& 13.60& 13.57& 13.61& 14.03&  20.82&     6.43&   1096.0&  230.0\\
22 33 02.70  -60 35 39.5 &         & 20.79& 18.98 &17.77& 17.22& 16.80& 16.67& 16.51& 16.86&   2.33&     7.42&         &       \\
22 33 02.76  -60 32 13.3 &  M3 V   & 23.00& 21.26 &19.76& 18.87& 17.74& 16.90& 16.76& 17.01&   1.38&     0.70&    105.0&   23.0\\
22 33 03.07  -60 32 30.8 &  M1 V   & 20.01& 18.15 &16.83& 16.20& 15.53& 14.97& 14.73& 15.04&   9.29&     4.28&    550.0&  118.0\\
22 33 08.20  -60 33 21.2 &  K4 V   &      &       &     & 17.30&      & 16.75& 16.57& 16.97&   1.91&    12.14&         &       \\
22 33 12.09  -60 34 16.7 &         &      &       &     &      &      & 22.46& 21.76& 21.10&   1.39&     8.16&         &       \\
22 33 15.83  -60 32 24.0 &  M2 V   &      & 15.80 &     & 13.95& 13.11& 12.47& 12.24& 12.53&  90.78&    31.03&   6000.0& 1311.0\\
22 33 19.00  -60 32 27.8 &  M2 V   &      &       &     & 20.83&      &      &      &      &   1.53&     6.40&         &       \\
22 33 20.83  -60 34 35.1 &  G1 V   &      &       &     & 17.10&      &      &      &      &       &     6.74&         &       \\
22 33 24.02  -60 33 10.6 &  G3.5 V &      &       &     & 17.91&      &      &      &      &   0.78&         &         &       \\
22 33 24.22  -60 33 52.9 &  M3 V   &      &       &     & 16.32&      &      &      &      &  16.07&    15.89&         &       \\
22 33 26.22  -60 32 05.9 &  F8 III &      &       &     & 15.93&      &      &      &      &       &         &         &       \\
22 33 28.05  -60 33 38.0 &  G5 V   &      &       &     & 19.96&      &      &      &      &       &         &         &       \\
22 33 28.93  -60 35 01.5 &  M2.5 V &      &       &     & 21.53&      &      &      &      &       &         &         &       \\
22 33 31.23  -60 33 43.9 &  M3 V   &      &       &     & 19.48&      &      &      &      &       &         &         &       \\
22 33 31.67  -60 33 41.9 &  M2 V   &      &       &     & 24.04&      &      &      &      &       &         &         &       \\
22 33 37.40  -60 34 03.2 &  M3 V   &      &       &     & 18.84&      &      &      &      &       &         &         &       \\
22 33 46.17  -60 34 03.4 &  M2 III &      &       &     & 19.95&      &      &      &      &       &         &         &       \\
\end{tabular}
\end{tiny}
\end{table*}

\begin{figure}
\epsfig{file=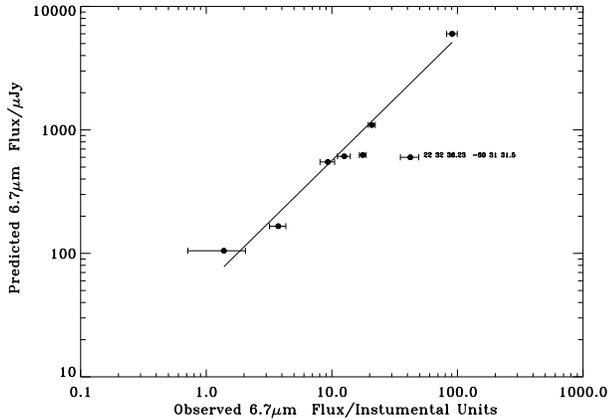,width=6cm,angle=90}
\caption{Calibration of the 6.7$\mu$m data.  The linear fit is
constrained to pass through the origin and excludes source
22~32~36.23~-60~31~31.5 .  The fit is $Sp/\mu{\rm Jy}=56\pm20So/
{\rm Instrumental Unit}$.}\label{fig:calib_7}
\end{figure}

\begin{figure}
\epsfig{file=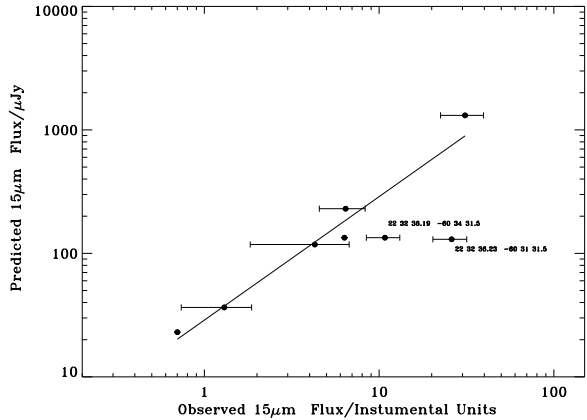,width=6cm,angle=90}
\caption{Calibration of the 15$\mu$m data.  The linear fit is
constrained to pass through the origin and excludes sources
22~32~36.23~-60~31~31.5 and 22~32~36.19~-60~34~31.5.  
The fit is $Sp/\mu{\rm Jy}=29\pm8So/{\rm Instrumental Unit}$.}\label{fig:calib_15}

\end{figure}

For each star we returned to the {\em ISO} maps and extracted the flux
and uncertainty in the flux, exactly as we did for our {\em ISO}
detected source list. For a number of them
which had good {\em ISO} detections at
6.7$\mu$m and good optical information we estimated the expected
6.7$\mu$m and 15$\mu$m fluxes using the models of Kurucz\footnote{see {\tt kurucz.harvard.edu/}}.  
Where a spectral classification was not available, the temperature was
estimated from the optical photometry alone.
The predicted fluxes are plotted against the observed fluxes for the
two wavebands in Figures \ref{fig:calib_7} \&  \ref{fig:calib_15}.  We
performed a linear fit to these data constrained to pass through the
origin.  For the 6.7$\mu$m data we excluded the 
star at 22~32~36.23~-60~31~31.5, which was a significant outlier and
had neither a spectral type, nor NIR magnitudes.  For the 15$\mu$m
data we excluded  22~32~36.23~-60~31~31.5, 
as we had at  6.7  $\mu$m, and  also  22~32~36.19~-60~34~31.5,
which was an outlier in both fits and also had neither NIR nor
classification information.  
The fits are remarkably good once the few outliers have been excluded,
and provide us with a good flux calibration, which is used throughout
this paper. From the scatter in the correlation the errors in this 
calibration are estimated to be 36 per
cent at 6.7$\mu$m and 28 per cent at 15$\mu$m.

\section{Simulations}\label{simulations}

The noise properties of the {\em ISO} data are sufficiently
complicated that simulations
are essential in order to determine accurately the quality of the
information extracted from our maps.  
In particular we wish to assess the reliability of the source
catalogues presented in Section \ref{catalogues} and to calculate the effective
area over which we could have detected sources above a given flux
limit, to facilitate computation of source counts in Section
\ref{counts}. We have thus constructed a number of simulated datasets which
mimic the noise properties of the real data as faithfully as
possible.

\subsection{Method}

As discussed earlier, most sources will not be detected significantly
in individual time-lines,  only appearing  after co-addition of many 
observations of the same patch of sky.  By corrupting the astrometric
information before constructing the maps it is thus possible to remove 
most of the real source signal, the remaining fluctuations being
almost entirely due to
noise.   This technique was employed in our reduction of the {HDF
North} data, where we corrupted the astrometric information by
randomizing the apparent location of each pixel within the detector array for each
pointing position.  This technique was
not entirely satisfactory, since noise that was correlated between
neighbouring pointings and pixels was also artificially reduced in the 
resulting maps.  A much better technique is to corrupt the astrometric
information coherently for the whole detector, so that real sources
are still dispersed, while maintaining the time-ordering of the data,
and the localization of pixel groups.
This
technique should account for all sources of instrumental noise.
 There are
seven possible ways of achieving this for a given raster, corresponding
to each independent reflection and rotation of the detector through 90 degrees.
Real sources that happen to lie near the axis of symmetry in some
pointings will be less corrupted than others, so we  expect the
resulting ``noise'' maps to have some fluctuations due to real sources
and they are thus pessimistic estimates of the noise.  
For each observation we generated all seven possible ``noise'' maps.
Since the HDF South field is observed four times with each band
there are thus $7^4\equiv2401$ different combinations of ``noise'' maps that we
can use to simulate a completed map.

One source of noise that is not included in these ``noise'' maps is
confusion noise due to real sources. To account for this we generated
artificial source lists generated from a number count distribution
which was a reasonable fit to a preliminary analysis of the total
source counts (i.e. including both stars and galaxies).  We generated
25 independent synthetic source lists and added each of these sources
list to a randomly selected set of ``noise'' maps, using the
empirical point spread function discussed in Section \ref{sources}. 
The sources were
placed on the maps at their nominal positions, thus we did not include 
any additional uncertainties in the field distortions or registration, which
will be taken into account by our use of an empirical point spread function.
The resulting maps were co-added and processed to produce source lists
in exactly the same fashion as the real data.

\subsection{Results}

\begin{figure}
\epsfig{file=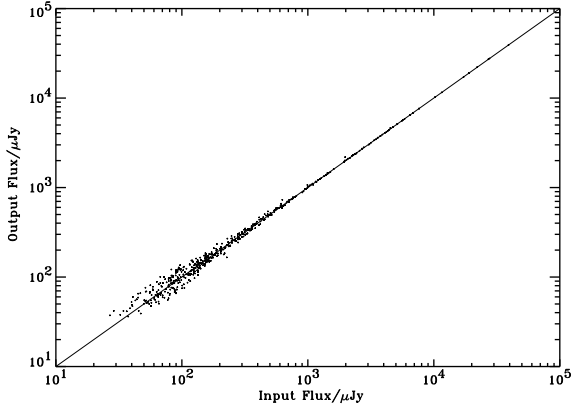,width=8cm}
\caption{Output flux vs Input flux from the simulated data at 6.7$\mu$m.
}\label{fig:in_out_7}
\end{figure}

\begin{figure}
\epsfig{file=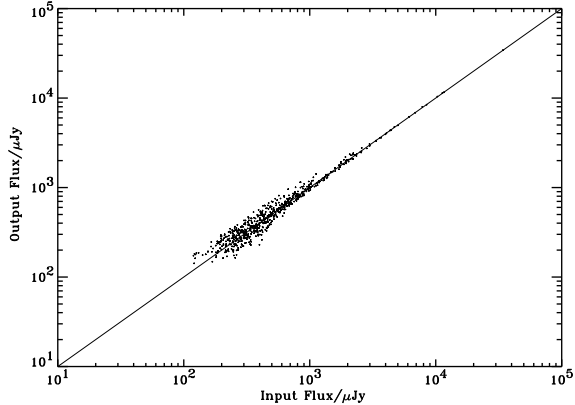,width=8cm}
\caption{Output flux vs Input flux from the simulated data at 15$\mu$m.
}\label{fig:in_out_15}
\end{figure}

Having extracted the sources from the simulated maps we can
immediately use these to check for any biases or non-linearities in
our flux estimation either from the peculiarities in the 
data reduction process or from the properties of the sky itself.
Biases from the sky might arise through confusion (where faint sources
are blured together and add flux to identified sources) or Eddington
bias (sometimes called Malmquist bias, where even with a symmetric
noise distribution, if the source counts are rising more sources are
randomly scattered to brighter fluxes than are randomly scattered to
faint fluxes).
The first step is to associate the detected
sources with the corresponding input source.   To do this we use a
4$\arcsec$ search radius and require that the ratio of the output and
input fluxes does not exceed $\pm0.3$ dex.

The resulting comparisons are shown in Figures ~\ref{fig:in_out_7}
\& \ref{fig:in_out_15}.   At 6.7$\mu$m the results are highly linear
and there appears to be no significant bias, while at 15$\mu$m, where we
expect the confusion noise to be higher, there is some evidence for a
small tendency to over-estimate faint fluxes.  For the purposes of this
paper we ignore these small flux biases.

\section{Catalogues}\label{catalogues}

The simulations of Section \ref{simulations} were used to investigate
how to construct a highly reliable source list that could be described 
simply in terms of the quantities introduced in Section \ref{sources}.
We eventually decided on the following criteria: 
\begin{enumerate}
\item $SNR0>3$: the initial candidate selection from the co-added maps
\item  all candidates located within a 2.5$\arcmin$ radius of  
22~32~56.2~-60~33~2.7 (J2000):  this excludes regions of low $NPOINT$, non-Gaussian
and/or  higher noise, giving a clean and simple selection
criterion
\item $SNR5>1$: this excluded spurious sources generated from a strong
noise feature in one map.
\item for LW2 only, $PICK>3$.
\end{enumerate}

These criteria resulted in 24 sources being detected in the LW2 maps
and an equal number being detected in the LW3 maps: 
from the simulations we estimate that the resulting source lists have
2.3 and 2.4 spurious sources, respectively, implying a
reliability of about 90 per cent.   The ``completeness'' of these
lists, expressed as an effective survey area, is discussed in Section \ref{counts}.

For ease of comparison of our results with those from any possible
future reductions of these data by other methods, we present our
source catalogues with fluxes given in both instrumental and physical
units, thereby decoupling issues of source detection, reliability and
completeness from that of photometric calibration. In Tables
\ref{tab:src7} and \ref{tab:src15} we list the separate source
catalogues for the two bands, while Table \ref{tab:merged} presents
the results of merging these two catalogues. 
This was done through associating
sources from different bands separated by less than 5 arcsec, which
produced thirteen matches.  The 5 arcsec radius was chosen to be
slightly less than the Airy radius at 15 $\mu$m (6 arcsec), larger
than the astrometric errors which are discussed further in Paper II,
yet small enough that there is little danger of association with 
an unrelated neighbouring source (the  number of unrelated sources
 expected in a circle this size is 0.03, assuming a Poission distribution).
For those sources where no such match was
found, we returned to the
smoothed signal--to--noise map and if we found a signal--to--noise ratio
above two we use the peak flux at this position (the assumption
being that the rejection criteria which were applied to ensure
reliability are not necessary since we have a confirmed detection
in the other wavelength). Otherwise we determine an upper-limit,
being the flux that would have given the observed smoothed signal
if the noise fluctuation was $-2\sigma$.
The absolute reference frame of the {\em ISO} data was set by making the
position of the brightest 15$\mu$m source, ISOHDFSC15~J223306-603350,
coincident with that of the bright radio source HDFS~J223306.0-603350
(A. Hopkins, {\em priv. comm.}) with which it is clearly associated, and
then optimising the match between the ISO and optical positions of several
of the bright stellar identifications of Paper II: this latter procedure
only shifted the astrometric frames of the 6.7 and 15$\mu$m data by 
$\sim$1 arcsec each.

\begin{table*}
\caption{Sources selected at 7$\mu$m. SNR is estimated from the
smoothed, co-added map.  Flux ($S$, in instrumental units) is
estimated from the peak in the
smoothed map, the error is estimated from the standard deviation of
the peak flux from independent rasters. $SNR5$ is $S/\sigma$.  $PICK$
is the number of independent $SNRn>1$ detections. }\label{tab:src7}
\begin{tabular}{lcrrrrr}
Name & RA \& Dec (J2000) & $SNR$ & $S$ & $\sigma$ & $SNR5$ & $PICK$ \\~ \\ \hline

  ISOHDFSC7~J223315-603224 &  22 33 15.75  -60 32 24.0 &  205.8 &  5083.7 &   492.8 &   10.0 & 4 \\
  ISOHDFSC7~J223259-603118 &  22 32 59.50  -60 31 18.9 &   66.6 &  1165.9 &    74.6 &   15.8 & 4 \\
  ISOHDFSC7~J223306-603349 &  22 33 06.08  -60 33 49.1 &   62.2 &   943.0 &     8.3 &  113.9 & 4 \\
  ISOHDFSC7~J223303-603230 &  22 33 03.04  -60 32 30.6 &   38.8 &   520.2 &    69.1 &    7.6 & 4 \\
  ISOHDFSC7~J223237-603256 &  22 32 37.50  -60 32 56.7 &   32.3 &   699.4 &    79.4 &    8.6 & 4 \\
  ISOHDFSC7~J223243-603351 &  22 32 43.63  -60 33 51.0 &   15.1 &   242.5 &    30.2 &    8.1 & 4 \\
  ISOHDFSC7~J223250-603359 &  22 32 50.58  -60 33 59.9 &   14.9 &   209.4 &    31.4 &    6.7 & 4 \\
  ISOHDFSC7~J223243-603242 &  22 32 43.03  -60 32 42.2 &   10.6 &   171.9 &    39.9 &    4.1 & 4 \\
  ISOHDFSC7~J223312-603350 &  22 33 12.36  -60 33 50.7 &   10.6 &   209.4 &    68.4 &    3.2 & 4 \\
  ISOHDFSC7~J223243-603441 &  22 32 43.57  -60 34 41.6 &    9.1 &   193.2 &    42.0 &    4.6 & 4 \\
  ISOHDFSC7~J223245-603418 &  22 32 45.53  -60 34 18.0 &    8.3 &   140.6 &    56.7 &    2.5 & 4 \\
  ISOHDFSC7~J223308-603317 &  22 33 08.13  -60 33 17.8 &    6.8 &   107.0 &    25.6 &    4.0 & 4 \\
  ISOHDFSC7~J223237-603235 &  22 32 37.99  -60 32 35.5 &    5.9 &   122.1 &    46.3 &    2.4 & 4 \\
  ISOHDFSC7~J223303-603336 &  22 33 03.43  -60 33 36.0 &    5.5 &    76.2 &    14.1 &    5.1 & 4 \\
  ISOHDFSC7~J223302-603213 &  22 33 02.62  -60 32 13.3 &    5.3 &    77.3 &    37.6 &    2.2 & 4 \\
  ISOHDFSC7~J223256-603059 &  22 32 56.79  -60 30 59.2 &    4.9 &    98.0 &    35.2 &    2.7 & 4 \\
  ISOHDFSC7~J223247-603336 &  22 32 47.81  -60 33 36.6 &    4.5 &    63.8 &    41.0 &    1.6 & 4 \\
  ISOHDFSC7~J223302-603323 &  22 33 02.76  -60 33 23.5 &    4.5 &    62.2 &    12.5 &    4.9 & 4 \\
  ISOHDFSC7~J223253-603328 &  22 32 53.07  -60 33 28.1 &    4.3 &    56.0 &    20.1 &    2.7 & 3 \\
  ISOHDFSC7~J223254-603115 &  22 32 54.95  -60 31 15.1 &    3.4 &    62.2 &    28.4 &    2.3 & 3 \\
  ISOHDFSC7~J223254-603127 &  22 32 54.55  -60 31 27.8 &    3.0 &    51.0 &    19.3 &    2.5 & 3 \\
  ISOHDFSC7~J223302-603137 &  22 33 02.12  -60 31 37.5 &    4.3 &    70.6 &    46.2 &    1.4 & 3 \\
  ISOHDFSC7~J223307-603247 &  22 33 07.53  -60 32 47.0 &    4.1 &    64.4 &    30.0 &    2.0 & 3 \\
  ISOHDFSC7~J223254-603143 &  22 32 54.92  -60 31 43.8 &    3.5 &    51.5 &    23.2 &    1.8 & 3 \\

\end{tabular}
\end{table*}

\begin{table*}
\caption{Sources selected at 15$\mu$m. SNR is estimated from the
smoothed, co-added map.  Flux ($S$, in instrumental units) is
estimated from the peak in the
smoothed map, the error is estimated from the standard deviation of
the peak flux from independent rasters. $SNR5$ is $S/\sigma$.  $PICK$
is the number of independent $SNRn>1$ detections. }\label{tab:src15}
\begin{tabular}{lcrrrrr}
Name & RA \& Dec (J2000) & $SNR$ & $S$ & $\sigma$ & $SNR5$ & $PICK$ \\~ \\ \hline

 ISOHDFSC15~J223306-603350 &  22 33 06.08  -60 33 50.0 &   50.8 &  2274.2 &    15.8 &  142.1 & 4 \\
 ISOHDFSC15~J223245-603418 &  22 32 45.72  -60 34 18.3 &   15.4 &   826.5 &    70.8 &   11.7 & 4 \\
 ISOHDFSC15~J223315-603223 &  22 33 15.81  -60 32 23.6 &   12.4 &   899.9 &   248.4 &    3.3 & 4 \\
 ISOHDFSC15~J223312-603349 &  22 33 12.25  -60 33 49.6 &   10.5 &   617.1 &    67.6 &    8.6 & 4 \\
 ISOHDFSC15~J223247-603335 &  22 32 47.63  -60 33 35.5 &    8.7 &   385.1 &   156.1 &    2.3 & 4 \\
 ISOHDFSC15~J223307-603248 &  22 33 07.54  -60 32 48.8 &    8.6 &   399.6 &    66.8 &    5.2 & 4 \\
 ISOHDFSC15~J223245-603226 &  22 32 45.81  -60 32 26.1 &    8.4 &   412.7 &    56.7 &    6.7 & 4 \\
 ISOHDFSC15~J223257-603305 &  22 32 57.42  -60 33 05.7 &    8.1 &   308.8 &   118.0 &    2.4 & 4 \\
 ISOHDFSC15~J223308-603314 &  22 33 08.01  -60 33 14.9 &    7.6 &   352.1 &    96.4 &    3.5 & 4 \\
 ISOHDFSC15~J223254-603129 &  22 32 54.49  -60 31 29.7 &    7.1 &   361.6 &    96.5 &    3.3 & 4 \\
 ISOHDFSC15~J223251-603335 &  22 32 51.81  -60 33 35.1 &    6.2 &   255.2 &    56.1 &    4.2 & 4 \\
 ISOHDFSC15~J223252-603327 &  22 32 52.87  -60 33 27.9 &    6.1 &   239.8 &    58.2 &    3.7 & 4 \\
 ISOHDFSC15~J223243-603351 &  22 32 43.42  -60 33 51.6 &    5.9 &   310.9 &     4.4 &   70.6 & 3 \\
 ISOHDFSC15~J223243-603440 &  22 32 43.44  -60 34 40.7 &    4.7 &   312.0 &   158.9 &    2.0 & 4 \\
 ISOHDFSC15~J223306-603436 &  22 33 06.05  -60 34 36.6 &    4.3 &   241.6 &    95.6 &    2.8 & 3 \\
 ISOHDFSC15~J223306-603450 &  22 33 06.99  -60 34 50.8 &    4.2 &   242.4 &    64.0 &    3.8 & 3 \\
 ISOHDFSC15~J223243-603243 &  22 32 43.12  -60 32 43.3 &    4.1 &   216.9 &    88.6 &    2.5 & 3 \\
 ISOHDFSC15~J223312-603416 &  22 33 12.25  -60 34 16.9 &    3.7 &   236.6 &    72.2 &    3.1 & 3 \\
 ISOHDFSC15~J223244-603455 &  22 32 44.22  -60 34 55.3 &    3.5 &   239.5 &    99.0 &    2.6 & 3 \\
 ISOHDFSC15~J223256-603513 &  22 32 56.46  -60 35 13.2 &    3.4 &   217.8 &    31.8 &    6.8 & 3 \\
 ISOHDFSC15~J223259-603116 &  22 32 59.90  -60 31 16.6 &    3.3 &   185.3 &    48.6 &    3.6 & 4 \\
 ISOHDFSC15~J223244-603110 &  22 32 44.70  -60 31 10.0 &    3.1 &   242.7 &    52.9 &    4.7 & 3 \\
 ISOHDFSC15~J223240-603141 &  22 32 40.53  -60 31 41.0 &    3.1 &   235.5 &    45.6 &    5.3 & 4 \\
 ISOHDFSC15~J223314-603203 &  22 33 14.40  -60 32 03.4 &    3.3 &   240.1 &   178.2 &    1.0 & 2 \\

\end{tabular}
\end{table*}

\begin{table*}
\caption{Merged {\em ISO} HDF South source list.  The $6.7\mu$m and
$15\mu$m source have been cross-correlated using a 5\arcsec search
radius, and fluxes for non-matches are determined from the maps.  
Upper-limits are also estimated from the maps and denoted by ``$<S_{\rm up}$'',
where $S_{\rm up}=S_{\rm obs}+2\sigma$, and  $\sigma$
is estimated from the scatter between independent
maps at the source position and $S_{\rm obs}$ is the recorded
flux at the source position (which may be negative).
Sources marked with an asterisk do not have optical identifications (see Paper II).
In most cases they are located near another bright source and
are very likely to be spurious: their number is  consistent with our simulations, 
and these
``sources'' have not been used in our number count analysis.}
\label{tab:merged}
\begin{tabular}{lrrrr}

Name	& $S_{6.7}$ & $\sigma_{6.7}$& $S_{15}$ & $\sigma_{15}$\\
	&$/\mu$Jy&$/\mu$Jy&$/\mu$Jy&$/\mu$Jy\\~\\ \hline
  ISOHDFS~J223237-603256&$   699.4 $&$    79.4 $&$  <389.9 $&$   168.4 $ \\
  ISOHDFS~J223237-603235&$   122.1 $&$    46.3 $&$  <321.1 $&$   122.5 $ \\
  ISOHDFS~J223240-603141&$  <119.4 $&$    43.7 $&$   235.5 $&$    45.6 $ \\
  ISOHDFS~J223243-603242&$   171.9 $&$    39.9 $&$   216.9 $&$    88.6 $ \\
  ISOHDFS~J223243-603441&$   193.2 $&$    42.0 $&$   312.0 $&$   158.9 $ \\
  ISOHDFS~J223243-603351&$   242.5 $&$    30.2 $&$   310.9 $&$     4.4 $ \\
  ISOHDFS~J223244-603455&$   <53.6 $&$    21.3 $&$   239.5 $&$    99.0 $ \\
  ISOHDFS~J223244-603110&$   <90.4 $&$    31.4 $&$   242.7 $&$    52.9 $ \\
  ISOHDFS~J223245-603418&$   140.6 $&$    56.7 $&$   826.5 $&$    70.8 $ \\
  ISOHDFS~J223245-603226&$   <46.7 $&$    19.3 $&$   412.7 $&$    56.7 $ \\
  ISOHDFS~J223247-603335&$    63.8 $&$    41.0 $&$   385.1 $&$   156.1 $ \\
  ISOHDFS~J223250-603359&$   209.4 $&$    31.4 $&$  <101.9 $&$    66.5 $ \\
  ISOHDFS~J223251-603335&$    31.6 $&$    28.9 $&$   255.2 $&$    56.1 $ \\
  ISOHDFS~J223252-603327&$    56.0 $&$    20.1 $&$   239.8 $&$    58.2 $ \\
  ISOHDFS~J223254-603129&$    51.0 $&$    19.3 $&$   361.6 $&$    96.5 $ \\
  ISOHDFS~J223254-603143&$    51.5 $&$    23.2 $&$   <53.9 $&$    20.2 $ \\
  ISOHDFS~J223254-603115&$    62.2 $&$    28.4 $&$   152.5 $&$    57.2 $ \\
  ISOHDFS~J223256-603513$^*$&$   <43.7 $&$    30.1 $&$   217.8 $&$    31.8 $ \\
  ISOHDFS~J223256-603059$^*$&$    98.0 $&$    35.2 $&$  <134.8 $&$    59.1 $ \\
  ISOHDFS~J223257-603305&$   <69.3 $&$    22.6 $&$   308.8 $&$   118.0 $ \\
  ISOHDFS~J223259-603118&$  1165.9 $&$    74.6 $&$   185.3 $&$    48.6 $ \\
  ISOHDFS~J223302-603137$^*$&$    70.6 $&$    46.2 $&$   <12.1 $&$     9.0 $ \\
  ISOHDFS~J223302-603213&$    77.3 $&$    37.6 $&$  <143.3 $&$    70.8 $ \\
  ISOHDFS~J223302-603323&$    62.2 $&$    12.5 $&$    82.5 $&$   106.0 $ \\
  ISOHDFS~J223303-603230&$   520.2 $&$    69.1 $&$   114.9 $&$    65.1 $ \\
  ISOHDFS~J223303-603336&$    76.2 $&$    14.1 $&$   <74.7 $&$    40.6 $ \\
  ISOHDFS~J223306-603436&$    49.8 $&$    19.8 $&$   241.6 $&$    95.6 $ \\
  ISOHDFS~J223306-603349&$   943.0 $&$     8.3 $&$  2274.2 $&$    15.8 $ \\
  ISOHDFS~J223306-603450&$   <87.4 $&$    41.3 $&$   242.4 $&$    64.0 $ \\
  ISOHDFS~J223307-603248&$    64.4 $&$    30.0 $&$   399.6 $&$    66.8 $ \\
  ISOHDFS~J223308-603314&$   107.0 $&$    25.6 $&$   352.1 $&$    96.4 $ \\
  ISOHDFS~J223312-603416&$    61.2 $&$    65.0 $&$   236.6 $&$    72.2 $ \\
  ISOHDFS~J223312-603350&$   209.4 $&$    68.4 $&$   617.1 $&$    67.6 $ \\
  ISOHDFS~J223314-603203&$   100.5 $&$    82.8 $&$   240.1 $&$   178.2 $ \\
  ISOHDFS~J223315-603224&$  5083.7 $&$   492.8 $&$   899.9 $&$   248.4 $ \\

\end{tabular}
\end{table*}

\begin{figure}
\epsfig{file=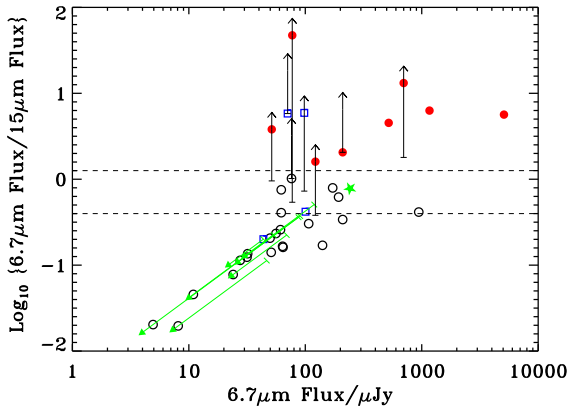,width=6cm,angle=90}
\caption{6.7$\mu$m/15$\mu$m colour as a function of 6.7$\mu$m flux.
Sources morphologically classified as stars are indicated with filled
symbols, while other sources have open symbols. 
ISOHDFS~J223243-603351, which has  a broad line in the optical
spectrum is indicated by 5--pointed star.  Objects with no reliable 
optical counterpart are indicated by squares.
Upper-limits
are indicated by  arrows, the tail of the arrow begins
at the position given by the upper-limit, if a positive flux
measurement was recorded the plotting symbol is placed at the
position inferred from this measurement, otherwise the plotting symbol
is placed at the tail of the arrow, which is then of arbitrary
length.  Upward-pointing arrows indicate upper-limits in the 15$\mu$m 
flux, arrows pointing to the bottom left indicate upper-limits in the 6.7$\mu$m flux.
}\label{fig:lw2_col_lw3}
\end{figure}

The resulting $S_{6.7}/S_{15}$ colour--flux diagram is shown  in Figure
\ref{fig:lw2_col_lw3}.  
Notice a reasonably clear distinction between
stars and galaxies.  For $\log_{10}(S_{6.7}/S_{15})>0.1$ all sources
with an identification are morphologically classified as stars and occupy, or
are consistent with, a well defined
stellar locus with $\log_{10}(S_{6.7}/S_{15}) \sim 0.7$.
Below  $\log_{10}(S_{6.7}/S_{15})<-0.3$ everything is morphologically
non-stellar.  In between these two limits are a mixture of sources.
Firstly there are three stars which have upper limits 
at 15$\mu$m within this 
part of the diagram but are consistent with the 
colours of the other stars. Secondly there is one source
(ISOHDFS~J223243-603351) which 
 has a broad line in its optical spectrum, and we thus assume that
the relatively warm {\em ISO} colours are because the
infrared emission arises from a dusty torus being heated by an AGN.  
This region and above also contains two of the three objects which had no reliable 
optical association (ISOHDFS~J223256-603059 and ISOHDFS~J223302-603137)
as discussed in Paper II, and a third (ISOHDFS~J223314-603203) which is one of the
most uncertain identifications made there: in that paper we
note that each of these is located close to a bright star which 
may mean that the sources are spurious, and will certainly 
affect their {\em ISO} colours.  Of the four galaxies located 
firmly within this region we show in Paper II
that three have SEDs
consistent with normal spiral or cirrus galaxies
(ISOHDFS~J223243-603441, ISOHDFS~J223302-603323 and
ISOHDFS~J223303-603336, the last of these has only
an upper-limit at 15$\mu$m) while the fourth ISOHDFS~J223243-603242
does appear to have anomalously low 15$\mu$m flux indicating perhaps
a deeper rest--frame 10$\mu$m absorption.  Finally, two galaxies
lie on the lower colour limit, one of these (ISOHDFS~J223306-603349) is a 
bright spiral galaxy with a normal-spiral SED, while the second
ISOHDFS~J223254-603115 is associated with one of a pair of possibly 
interacting galaxies: this confuses the optical magnitudes is also
likely to confuse the {\em ISO} colours. 
We also note that
ISOHDFS~J223256-603513, for which we are unable to make a secure
optical
identification, was the only source for which we recorded a 
negative 7$\mu$m flux measurement when determining the upper-limits.

In conclusion, it appears that the  7/15$\mu$m flux ratio separates
quite neatly stars from star--forming galaxies, with AGN and normal
galaxies occupying a middle ground.  The {\em ISO} colours appear
consistent with our optical morphological classifications: for the 
purposes of this paper we will use the morphological
classifications and for the number count analysis will exclude
the unidentified sources.

\section{Source Counts}\label{counts}

\subsection{Calculation of effective area}

\begin{figure}
\epsfig{file=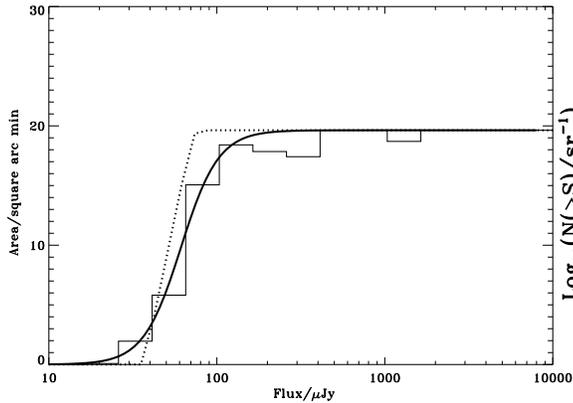,width=8cm}
\caption{6.7$\mu$m effective area as a function of flux.   The
histogram is estimated directly from fraction of sources detected from
the simulated data as a function of input source flux.  The dotted
line is a na\"{\i}ve estimate based simply on the SNR selection criterion
and the noise maps, while the solid line is a fit to the simulated histogram:
$\Omega=19.63 \frac{1}{2}\tanh [4.47 \log_{10} (S/61.1\mu{\rm Jy})+1]$
square arc min.
}\label{fig:lw2_eff_area}
\end{figure}

\begin{figure}
\epsfig{file=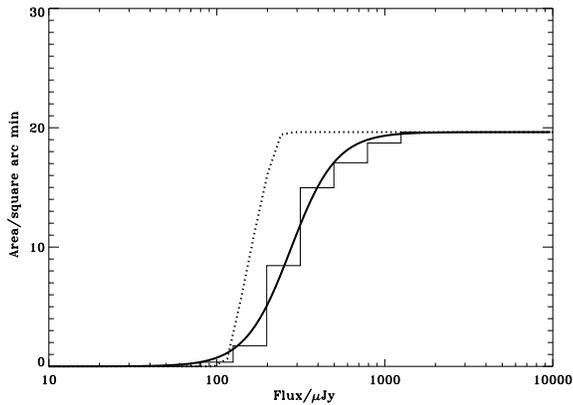,width=8cm}
\caption{15$\mu$m effective area as a function of flux.   The
histogram is estimated directly from fraction of sources detected from
the simulated data as a function of input source flux.  The dotted
line is a na\"{\i}ve estimate based simply on the SNR selection criterion
and the noise maps. while the solid line is a fit to the simulated histogram:
$\Omega=19.63 \frac{1}{2}\tanh [3.70 \log_{10} (S/275\mu{\rm Jy})+1]$
square arc min.
}\label{fig:lw3_eff_area}
\end{figure}

From the simulated catalogues we can directly determine the
``completeness'', which we define in terms of the effective area as a
function of flux; the effective area of the catalogue, at a given flux,  
being the area within which a
source of that flux could have been detected.   We can estimate this from the
simulations by determining the fraction of sources of a given input
flux which pass all our selection criteria, and 
the effective area is then this fraction of the area over which the input
catalogues were prepared: in practice we examine only those input
source which fell within a 2.5\arcmin radius of  
22~32~56.2~-60~33~2.7~(J2000).
If the output flux is an unbiased estimator of the input flux this
estimate of the effective area is unbiased.  Eddington bias and
confusion noise do cause some bias, as discussed in Section
\ref{simulations}, but these appear to be small and are only second
order effects in the calculation of effective area so we ignore them
here.  The resulting effective areas are illustrated in Figures
\ref{fig:lw2_eff_area} \& \ref{fig:lw3_eff_area}.

As with the ELAIS 6.7 and 15$\mu$m counts (Serjeant~et~al. 2000)
we fit the histogram of the effective area with a hyperbolic tan
function $\Omega=19.63 \frac{1}{2}\tanh [a \log_{10} (S/b)+1]$ where
$a$ defines the gradient of the decline and $b$ defines its location.
For the $6.7\mu$m simulations we find $a=4.47,b=61.1\mu{\rm Jy}$ while
for the $15\mu$m simulations we find $a=3.70,b=275\mu{\rm Jy}$.  Also
illustrated in Figures \ref{fig:lw2_eff_area} \&
\ref{fig:lw3_eff_area} are estimates of the effective area na\"{\i}vely
based on the noise maps used in constructing the $SNR$ maps.  Since
this ignores some of the selection criteria it over-predicts the
fraction of sources detectable at brighter fluxes. On the other hand
it takes no account of the boosting of faint input fluxes by confusion
noise or Eddington bias and so under-predicts the effective area at
faint fluxes. This illustrates that these simulations at least account
for these biases to the first order.

\subsection{Source Count Results}

\begin{figure}
\hspace{-1.6cm}
\epsfig{file=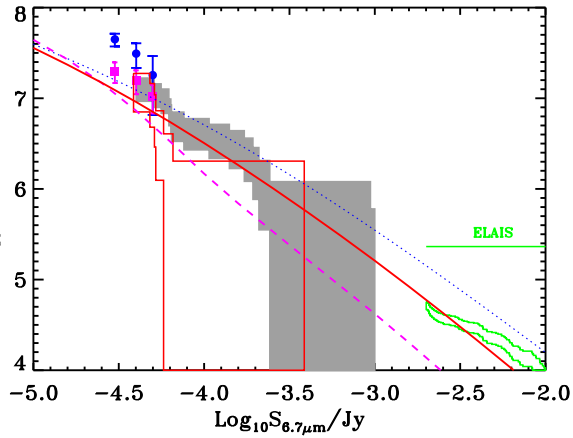,width=9cm}
\caption{Extragalactic counts at 6.7$\mu$m from HDF South in this
work (shaded region). Counts from other surveys are
illustrated as follows: 
Abell 2390  (Altieri et al. 1999, filled circles); 
Lockman Hole (Taniguchi et al. 1997, filled squares);
HDF North (Oliver~et~al.~ 1997, open region at faint fluxes); 
and ELAIS  (Serjeant~et~al.~2000, open region). 
The models of Rowan-Robinson (2001) (solid);
Pearson \& Rowan-Robinson (1996) (dotted) and  
Franceschini et al.~(1994) (short dashed) 
are plotted for comparison.}
\label{fig:counts_7}
\end{figure}

\begin{figure}
\hspace{-1.6cm}
\epsfig{file=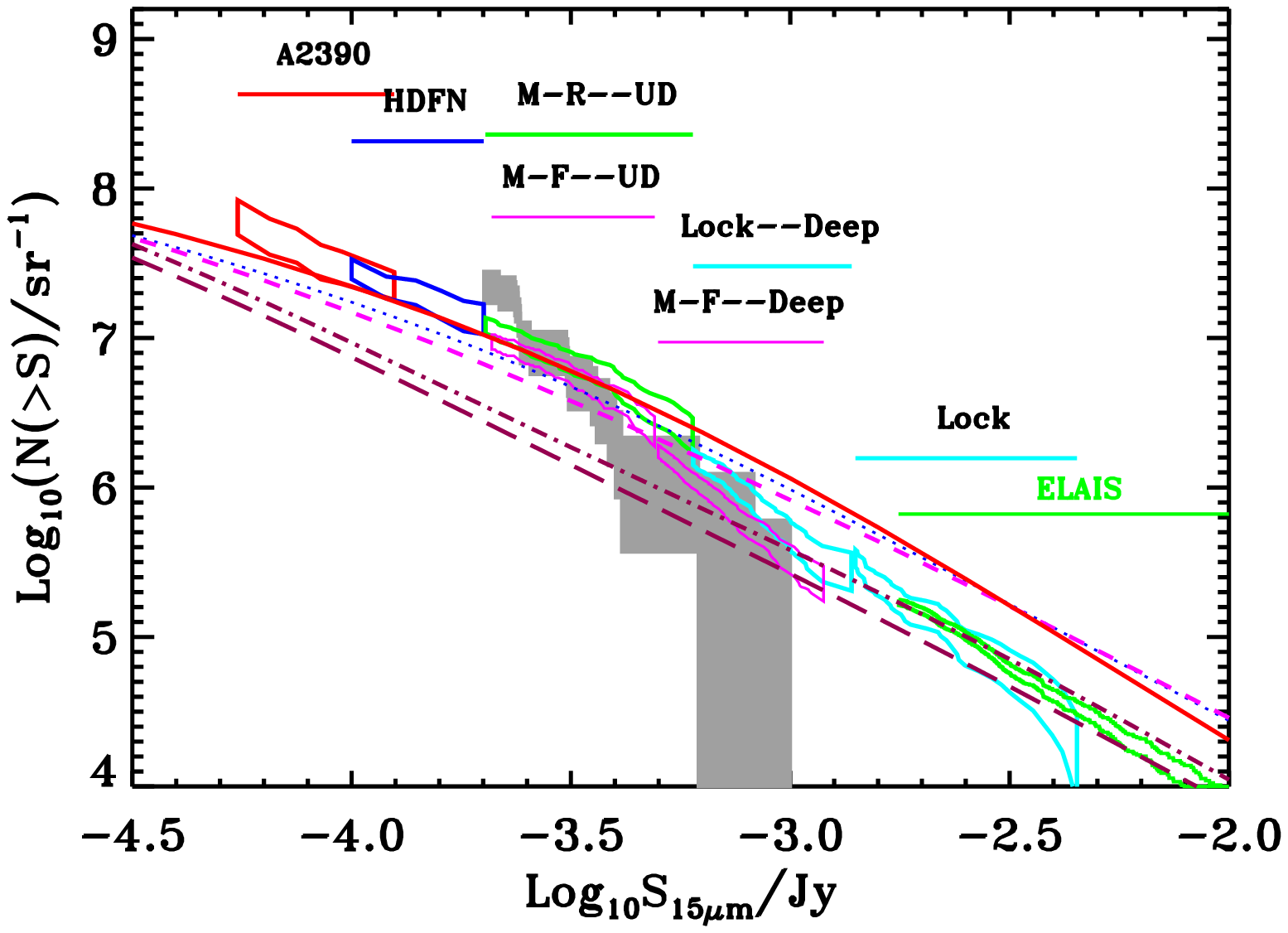,width=9cm}
\caption{Extragalactic source counts at 15$\mu$m with the counts
from HDF South in this
work shown as a shaded region. Counts from other surveys are
illustrated with open boxes and annoted above, in decreasing sensitivity:
Abell 2390 \protect\cite{Altieri et al. 1999}; 
Deep HDF North  \protect\cite{Aussel et al. 1999}; 
Marano-ROSAT Ultra-deep   \protect\cite{Elbaz et al. 1999};
Marano-Firback Ultra-deep (thin) \protect\cite{Elbaz et al. 1999};
Marano-Firback Deep (thin) \protect\cite{Elbaz et al. 1999};
Lockmann Deep \protect\cite{Elbaz et al. 1999};
Lockmann Shallow  \protect\cite{Elbaz et al. 1999};
ELAIS  \protect\cite{Serjeant et al. 2000}.
The models of Rowan-Robinson (2001) (solid);
Pearson \& Rowan-Robinson (1996) (dotted);  
Franceschini et al.~(1994) (short dashed); 
Guiderdoni et al. (1998) model 'A' (long dashed);
and Guiderdoni et al. (1998) model 'E' (dot--dash)
are plotted for comparison.}
\label{fig:counts_15}
\end{figure}

In Figures~\ref{fig:counts_7} and ~\ref{fig:counts_15} we plot the
resulting integral counts for galaxies (as defined by the colour
criterion defined in Section \ref{catalogues}).  For comparison we also
show the counts obtained by Oliver et al. (1997) for the HDF North.  In
Figure~\ref{fig:counts_15} we also show the counts derived by Aussel
et al. (1999) from the same HDF North data.

At 6.7$\mu$m the galaxy counts from HDF North and South are in
good agreement at the faint end, but the HDF South counts are
higher at the brighter end. One possible explanation for this is that
original selection of the HDF North field, which avoided bright
galaxies, biased the bright counts downwards.
 Due to the observational strategies
discussed in Section \ref{obs}, the HDF South data at 6.7$\mu$m
are  considerably superior to the data from the North.   The larger areal
coverage in particular means the brighter counts are better
constrained.   So, while Oliver~et~al.~(1997) suggested that the Pearson \&
Rowan-Robinson (1996) model could be ruled out, as it over-predicted the
number of bright sources, we find that HDF South data are much
more consistent with the Pearson  \&
Rowan-Robinson model;  indeed the Franceschini et al. (1994) model now
appears to be ruled out as it does not predict enough brighter
$6.7\mu$m sources.  Such a conclusion is also borne out by the
ELAIS counts (Serjeant et al. 2000) which are shown for comparison.

The 15$\mu$m HDF South data  over the range $250-400\mu$Jy are in striking agreement with Aussel et
al. (1999) HDF North counts (which push the {\em ISO data} deeper than those of
Oliver et al. 1997). If not coincidental, this agreement would
suggest that these populations are at moderate--to--high redshifts.  If
all the sources had been located at low redshift the volume sampled within
either HDF area ($\sim 20$ square arcmin) would be small and the
fluctuations due to cosmic variance large: {\em e.g.} if the sample were
limited to $z<0.5$, the fluctuations in this volume would be $>100$ per
cent (assuming a cubical geometry and the power spectrum of Peacock
and Dodds 1994), while, even at $z<1$ the fluctuations would be around 80 per
cent (e.g. Oliver et al. 2000).
Below $250\mu$Jy the HDF
South counts take  a sharp up-turn.  This flux level is also where the 
effective area drops,  and where we are susceptible to errors in its
calculation.  
We demonstrate in Paper II that the many of the sources appear to have
modest redshifts and so the agreement between the HDF North and South
brighter data may be coincidental, and the steep upturn could also
be an effect of clustering.
It would be possible to combine the HDF North and
South counts to produce a single determination of the counts over the
$250-400\mu$Jy regime, reducing the statistical errors by a factor of 
$\sqrt{2}$.  Given the uncertainty in the effective area below
$250\mu$m both the Pearson \& Rowan-Robinson (1996) model appear to provide
acceptable fits to the counts.  Both Guiderdoni et al. (1998) models
appear to be too far below the counts.

\begin{figure}
\epsfig{file=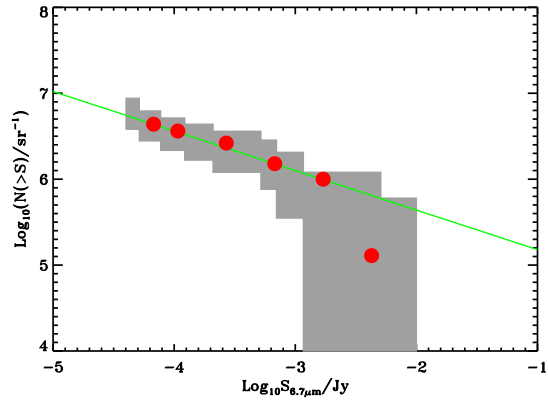,width=8cm}
\caption{The star source counts at 6.7$\mu$m in the  HDF South,
shaded region.
Circles indicate K--band star counts in the SGP transformed from 
\protect\cite{Minezaki et al. 1998} and the line  indicates a
fit to those data (excluding the brightest point)
}
\label{fig:star_counts_7}
\end{figure}

The star counts at $6.7\mu$m are shown in Figure \ref{fig:star_counts_7}.
There are currently no published star count data at this wavelength
and  flux to compare these data with.  However, it is possible to
extrapolate from star counts at shorter wavelengths. 
Minezaki et al. (1998) present near infrared (K--band) star
counts, towards the South Galactic Pole ($l=316, b=-89$).  
Converting  their differential counts into integral counts and 
using the K--band zero point of 673Jy we are able to compare their
data with ours. With the exception of their brightest point, their data
are fit with a power law $\log{N(>S)}=4.72-0.46 \cdot S_{K}$.  By coincidence,
the same power law is consistent with our data.  This is consistent 
with our detecting the same populations, since the slope is the same,
but with a higher number density (as we would expect our stars
to be fainter).
Since we would expect to be in the Rayleigh-Jeans part of a
Planck spectrum, we would expect our stars to fainter by a factor of
around 11, so the coincidence in normalization thus implies that
the Hubble Deep Field South has about three times the number density of
stars as the South Galactic Pole.

\section{Conclusions}

We have performed a survey using {\em ISO-CAM} at 6.7 and 15 $\mu$m in
the Hubble Deep Field South region. The observational and data
reduction techniques that we have employed mean that these data, and
the 6.7$\mu$m data in particular, are significantly improved over the
equivalent {\em ISO-HDF} data.  From the resulting data we have
extracted conservative bright source lists.  We have throughly
investigated the completeness and reliability of these lists using
simulations.  We have performed an external calibration of the data
using stars in the field, a number of which have been
spectroscopically classified.  We find that the 6.7 and 15$\mu$m
colour flux diagram provides a useful discriminant between stars and
galaxies.  We have investigated the number counts of the
extra-galactic sources and stars.  We find that the number counts of
the extra-galactic sources are consistent with previous
determinations, however, we stress that the volume sampled by our
survey is likely to small and so clustering effects (cosmic variance)
may mean that this agreement is somewhat coincidental.  A steep upturn
at the faintest fluxes is due to only a few sources and is almost
certain to be an effect of clustering. Further details of this project
can be found at {\tt astro.ic.ac.uk/hdfs}).

\section*{Acknowledgments}
We would like to thank the anonymous referee for rapid and
constructive feedback.
This paper is based on observations with {\em ISO}, an ESA project, with
instruments funded by ESA Member States (especially the PI countries:
France, Germany, the Netherlands and the United Kingdom) and with
participation of ISAS and NASA.
This work was in part supported by PPARC grant no.  GR/K98728
and  EC Network is FMRX-CT96-0068. We thank the ATNF HDF--S team, particularly Andrew Hopkins,
for providing us with radio data for our sources in advance of publication.

\label{lastpage}

\end{document}